\documentclass[twoside]{article}
\usepackage{fleqn,espcrc2}
\usepackage{graphicx}
\usepackage{epsfig}
\usepackage{amssymb,amsmath,array}

\title{Geometric Scaling at RHIC and LHC}

\author{Dani\"el Boer\address{Department of Physics and Astronomy,
    Vrije Universiteit Amsterdam, \\
    De Boelelaan 1081, 1081 HV Amsterdam, The Netherlands}
  and 
  Andre Utermann\address{Institut f\"ur Theoretische Physik,
    Universit\"at Regensburg, \\
    Universit\"atsstrasse 31, D-93040 Regensburg, Germany}
  and
  \underline{Erik Wessels}$^{\rm a}$}

\begin{document}

\begin{abstract}
  We demonstrate that the RHIC data for hadron production in $d$-$Au$
  collisions for all available rapidities are compatible with
  geometric scaling. In order to establish the presence of scaling
  violations expected from small-$x$ evolution a much larger range in
  transverse momentum and rapidity needs to be probed. We show that
  the fall-off of the transverse momentum distribution of produced
  hadrons at LHC is a sensitive probe of small-$x$ evolution.
\end{abstract}

\maketitle

%%%%%%%%%%%%%%%%%%%%%%%%%%%%%%%%%%%%%%%%%%%%%%%%%%%%%%%%%%%%%%%%%%%%%%%

It is well known that the small-$x$ DIS data show geometric scaling
\cite{Stasto:2000er}. This means that the cross section depends on
$Q^2/Q^2_s(x)$ only, instead of $Q^2$ and $x$ independently, where
$Q_s(x)$ is known as the saturation scale. Since geometric scaling
arises as a feature of saturation from nonlinear evolution equations,
such as the BK equation \cite{BK}, its occurrence is often seen as an
indication of saturation. Here, we investigate this issue by studying
the scaling properties of hadron collision data from RHIC at similarly
small $x$.

The small-$x$ inclusive HERA data were shown to be well described by
the phenomenological Golec-Biernat W\"usthoff (GBW) model for the
dipole cross section \cite{GBW}. In the GBW model, the cross section
is given by $\sigma = \sigma_0 N_{\rm GBW}(r_t,x)$, where $\sigma_0
\simeq 23\;{\rm mb}$ and the scattering amplitude $N_{\rm GBW}$ is
given by
\begin{equation}
N_{\rm GBW}(r_t,x) = 1-\exp\left(-\tfrac{1}{4} r_t^2
Q_s^2(x) \right), 
\label{NGBW}
\end{equation}
where $r_t$ is the dipole size. This amplitude depends on
$r^2_tQ^2_s(x)$ only, leading to a $Q^2/Q^2_s(x)$ dependence of the
DIS cross section. Hence, the amplitude (\ref{NGBW}) incorporates
geometric scaling. The saturation scale $Q_s(x)$ is parameterized as
\begin{equation}
Q_s(x) = Q_0\,
\left(\frac{x_0}{x}\right)^{\lambda/2}, \,\quad Q_0=1\,\mathrm{GeV}\,
\label{Qsx2}
\end{equation}
where $x_0 \simeq 3 \times 10^{-4}$ and $\lambda \simeq 0.3$
\cite{Stasto:2000er}.

To investigate whether saturation may be the cause of the observed
scaling, one can study geometric scaling in other experiments where
similarly small values of $x$ are probed, like $d\,$-$Au$ scattering
at RHIC. In terms of the dipole amplitude, the cross section of hadron
production in high-energy nucleon-nucleus collisions is given by
\cite{DHJ1,DHJ2}
\begin{eqnarray}
&&\hspace{-20pt}
\frac{ dN_h}{d\,y_hd^2p_t} = 
\frac{K(y_h)}{(2\pi)^2} \int_{x_F}^{1} dx_1 {\frac{x_1}{x_F}}  
\nonumber\\
&&\hspace{-20pt} \times
\bigg[\sum_q f_{q/p}(x_1,p_t^2) N_F \left(\tfrac{x_1}{x_F}p_t,x_2\right)
D_{h/q}\left(\tfrac{x_F}{x_1},p_t^2\right)
\nonumber \\
&&\hspace{-20pt}+
f_{g/p}(x_1,p_t^2) N_A \left(\tfrac{x_1}{x_F}p_t,x_2\right) 
D_{h/g}\left(\tfrac{x_F}{x_1},p_t^2\right)\bigg].
\label{eq:conv2}
\end{eqnarray}
Here $N_F$ describes a quark scattering off the small-$x$ field of the
nucleus, while $N_A$ applies to a gluon. The parton distribution
functions $f_{q/p}$ and the fragmentation functions $D_{h/q}$ are
taken at the scale $Q^2=p_t^2$, which we will always take to be larger
than 1 GeV$^2$.  The momentum fraction of the target partons equals
$x_2=x_1\exp(-2y_h)$. To good approximation one can neglect finite
mass effects, i.e.\ equate the pseudo-rapidity $\eta$ and the rapidity
$y_h$ and use $x_F=\sqrt{p_t^2+m^2}/\sqrt{s}\exp(\eta)\approx
p_t/\sqrt{s}\exp(y_h)$. Finally, there is an overall $y_h$-dependent
$K$-factor that effectively accounts for NLO corrections. The
$K$-factors are close to 1 in the forward region and become relevant
towards mid-rapidity.

Unlike in DIS, in hadron-hadron collisions, a scaling dipole amplitude
will not lead to a scaling property of the cross section, due to
convolutions with the parton distributions and fragmentation functions
in (\ref{eq:conv2}). Hence one cannot establish geometric scaling in
the data directly, but instead one has to test the scaling properties
of the dipole amplitude using a model, like the GBW model
(\ref{NGBW}). However, due to its exponential fall-off at large
transverse momentum, the GBW model cannot describe the RHIC data.
Instead, a modification of the GBW model, which we will refer to as
the DHJ model, was proposed in \cite{DHJ1,DHJ2}. It offers a good
description of hadron production in the forward region\footnote{As it
  turns out the study of Ref.\ \cite{DHJ2} contained an artificial
  upper limit on the $x_1$ integration to exclude large $x_2$. Without
  this cut, the larger $p_t$ data for $y_h=0, 1$ are not
  well-described by the DHJ model.}, and is given by
\begin{eqnarray} 
N_A({q}_t,x_2) &\equiv& \int d^2 r_t\: e^{i \vec{q}_t
    \cdot \vec{r}_t} \nonumber\\
&&\hspace{-50pt} \times
\left[1-\exp\left(-\tfrac{1}{4}(r_t^2
      Q_s^2(x_2))^{\gamma(q_t,x_2)}\right) \right]~.
\label{NA_param}
\end{eqnarray}
The corresponding expression $N_F$ for quarks is obtained from $N_A$
by the replacement $(r_t^2Q_s^2)^\gamma \to ((C_F/C_A)
r_t^2Q_s^2)^\gamma$, with $C_F/C_A=4/9$. Note that the so-called
``anomalous dimension'' $\gamma$ is a function of $q_t$ rather than
$r_t$, so that the Fourier transform can be obtained more easily.

The anomalous dimension $\gamma$ of the DHJ model is parameterized as
\begin{eqnarray}
\gamma_{\rm DHJ}(q_t,x_2) &=&\gamma_s\nonumber\\
&&\hspace{-65pt}
+(1-\gamma_s)
 \frac{|\log(q_t^2/Q_s^2(x_2))|}{\lambda
y+d\sqrt{y}+|\log(q_t^2/Q_s^2(x_2))|}. \label{gammaparam}
\end{eqnarray}
Here $y=\log 1/x_2$ is minus the rapidity of the target partons. The
saturation scale $Q_s(x_2)$ and $\lambda$ are taken from the GBW model
(\ref{Qsx2}), including a larger value of $Q_0\approx 1.63\;{\rm GeV}$
to account for the size of the nucleus. The parameter $d=1.2$ was
fitted to the data. Away from $Q_s$, $\gamma_{\rm DHJ}$ rises towards
1 like $(\log q_t/Q_s)/y$; clearly the model violates geometric
scaling. At the saturation scale $\gamma_{\rm DHJ}$ assumes a specific
value $\gamma_s=0.628$. The logarithmic rise, the scaling violations
and the value of $\gamma_s$ are expected from small-$x$ evolution
\cite{MTIIM}. However, an analysis of the BK equation suggests that a
smaller value of $\gamma\approx 0.44$ may be more appropriate at $Q_s$
\cite{BUW}. We note that the DHJ model (\ref{gammaparam}) was meant to
apply only outside the saturation region, which is hardly probed at
RHIC.  Hence the behaviour of (\ref{gammaparam}) for $q_t<Q_s$ is
irrelevant.

To investigate whether the RHIC data are compatible with geometric
scaling, we write down a new model that features exact scaling, unlike
the DHJ model. Further, the new model, which we will refer to as the
GS model, does not have the logarithmic rise or the value of $\gamma$
at $Q_s$ that are both expected from small-$x$ evolution.  Instead, it
is parameterized as
\begin{equation}
 \gamma_{\rm GS}(w=\tfrac{q_t}{Q_s})
 =\gamma_1+(1-\gamma_1)\frac{(w^a-1)}{(w^a-1)+b}\,.
\label{gamma_alpha}
\end{equation}
The parameters $a$ and $b$ will be fitted to the data. Not only is
$\gamma_{\rm GS}$ exactly scaling, it also rises much faster towards 1
at large $q_t$ than $\gamma_{\rm DHJ}$. Expanding the exponential, we
see that the large $q_t$ behavior of Eq.'s (\ref{gammaparam}) and
(\ref{gamma_alpha}) is given by \cite{buw2},
\begin{equation}
  N_A({q}_t) \stackrel{q_t \gg Q_s}{\propto}
\left\{\begin{array}{cl}
    \frac{Q_s^2}{q_t^4\log(q_t^2/Q_s^2)}& \hspace{0pt} 
\text{for $\gamma_{\rm DHJ}$}
\\[2ex]
\frac{Q_s^{2+a}}{q_t^{4+a}}& \hspace{0pt} \text{for $\gamma_{\rm GS}\,.$} 
\end{array}\right.
\label{NA_asympt}
\end{equation}
Hence, the $p_t$ distribution resulting from Eq.\ (\ref{gamma_alpha})
will fall-off much faster than in the DHJ model. We note that our
model (\ref{gamma_alpha}) is not intended to replace other,
theoretically better motivated models but is constructed simply to
investigate in a general way which conclusions can really be drawn
from the RHIC data in the central and forward regions.

\begin{figure}[htb]
\centering
\vspace{-15pt}
\includegraphics*[width=75mm]{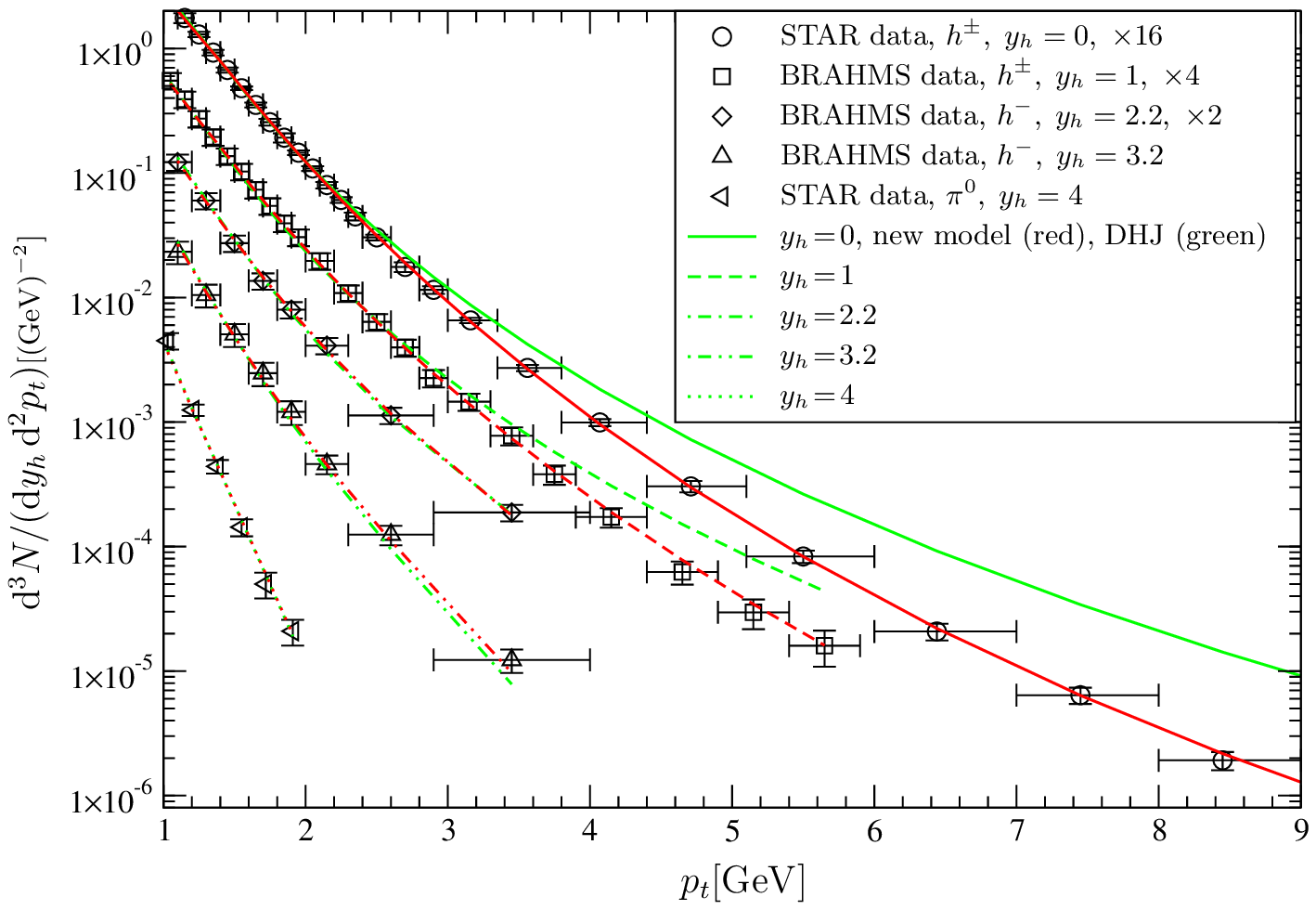}
\vspace{-35pt}
\caption{\label{fig_RHIC} Transverse momentum distribution of produced
  hadrons in $d$-$Au$ collisions as measured at RHIC (symbols) for
  various rapidities $y_h$. To make the plot clearer, the data and the
  curves for $y_h=0, 1$ and $2.2$ are multiplied with arbitrary
  factors, namely 16, 4 and 2, respectively.}
\end{figure}

In Fig.~\ref{fig_RHIC} we show $dN_h/(dy_h d^2p_t)$ following from
Eq.\ (\ref{eq:conv2}), calculated using the anomalous dimension
$\gamma_{\rm GS}$ in combination with the amplitude (\ref{NA_param}).
All $p_t$ distributions of produced hadrons measured at RHIC in
$d$-$Au$ collisions are well described. At $Q_s$ we have chosen for
\mbox{$\gamma_{\rm GS}(w=1)=\gamma_1$} the same value $\gamma_s=0.628$
as in the DHJ model, for ease of comparison. We also use the same
parameterization of $Q_s(x)$.  We obtain the best fit of the data for
$a=2.82$ and $b=168$.  As mentioned, this LO analysis requires a
$K$-factor to account for NLO corrections, which are expected to
become more relevant towards central rapidity.  The $p_t$-independent
$K$ factors we obtain for $y_h=0,1,2.2,3.2,4$ are given by $K=3.4,
2.9, 2.0,1.6, 0.7$ for our model, and $K=4.3, 3.3, 2.3, 1.7, 0.7$ for
the DHJ model. For further details of the calculation see \cite{buw2}.

%%%%%%%%%%%%%%%%%%
%%%%%%%%%%%%%%%%%%%%

We can conclude that the RHIC data are compatible with geometric
scaling at all rapidities. Hence, no scaling violations can be claimed
to be observed at RHIC. Of course, from this analysis such violations
cannot be ruled out either. The fact that in the forward region where
$y_h=2.2-4$ both the exactly scaling GS model and the scaling
violating DHJ model describe the data indicates that to reach a
conclusion on geometric scaling, a larger $\sqrt{s}$ is needed so that
a larger range of rapidities and $p_t$ can be probed. Further,
the logarithmic rise of $\gamma_{\rm DHJ}$ is ruled out in the central
region $y_h=0,1$.  However, the DHJ model breaks down only at
mid-rapidity and $p_t\gtrsim2.5$ GeV, where the values of $x_2$ become
larger than 0.01. Even though small-$x$ evolution may not be expected
to be valid anymore at $x\gtrsim0.01$, we note that in $d\,$-$Au$
collisions at RHIC $Q_s$ is still larger than in DIS at $x=0.01$.

\begin{figure}[htb]
\centering
\includegraphics*[width=75mm]{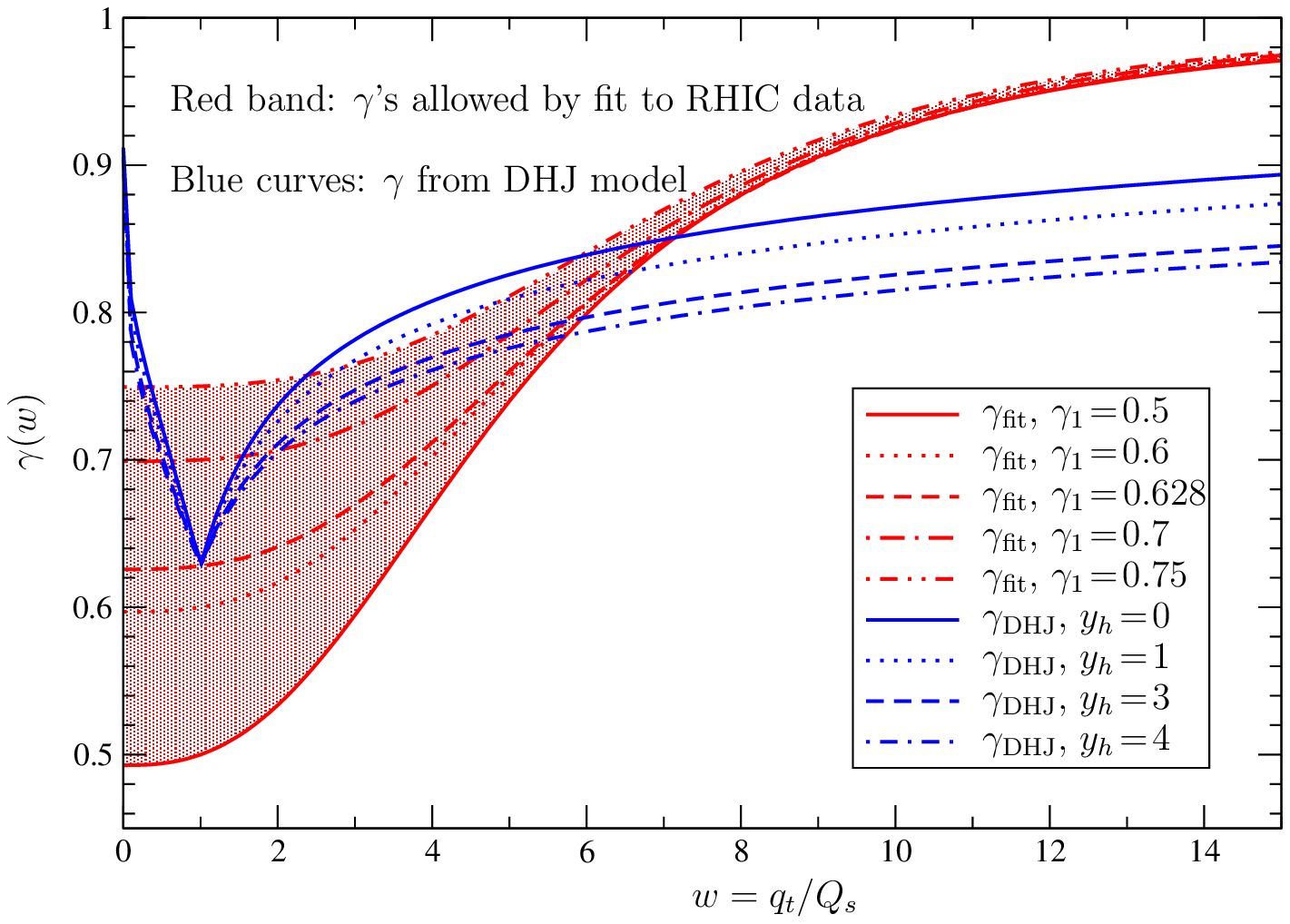}
\vspace{-35pt}
\caption{\label{fig_gam} Various fits of $\gamma_{\rm GS}(w)$, which
  describe the RHIC data equally well. For comparison we show curves
  representing $\gamma_{\rm DHJ}(w,y(w,y_h))$ at different rapidities
  $y_h$.}
\end{figure}

As mentioned, in the forward region both models (\ref{gammaparam}) and
(\ref{gamma_alpha}) work, although they have very different
properties.  To illustrate how sensitive the data are to the behavior
of $\gamma$, Fig.~\ref{fig_gam} shows various $\gamma_{\rm GS}(w)$'s
that describe the available data equally well. All are of the form
(\ref{gamma_alpha}), with different values of $\gamma_1$, $a$ and $b$.
The parameters that define the edges of the allowed region are
$\gamma_1=0.50$, $a=2.60$ and $b=70.2$ for the upper curve, and
$\gamma_1=0.75$, $a=3.10$ and $b=451$ for the lower curve.
Furthermore, we add lines representing $\gamma_{\rm DHJ}$
(\ref{gammaparam}) for different rapidities. To do so the rapidity of
the target parton $y$ needs to be expressed in terms of $w$ and $y_h$,
see \cite{buw2} for details. As said, the region below $Q_s$, where
the parameterization of $\gamma_{\rm DHJ}$ is not smooth, is hardly
probed at RHIC.  Fig.~\ref{fig_gam} shows that $\gamma$ is much more
constrained by the data at large $w$ compared to the region close to
$Q_s$. The reason for this is that around the saturation scale
$r_t=1/Q_s$, the integrand in the dipole scattering amplitude
(\ref{NA_param}) depends only weakly on $\gamma$. In addition, the
forward data, $y_h=3.2$ and 4, are essentially sensitive only to
$\gamma_1$, since for kinematic reasons they probe the region where
$w$ is close to 1.  Therefore, the rise of $\gamma$ with $w$ is
effectively constrained only by the data for $y_h=0,1$. From
Fig.~\ref{fig_gam} we see that $\gamma_{\rm DHJ}$ falls outside the
band of allowed $\gamma$'s at large $w$, but remains inside the band
for $w\lesssim2$---this corresponds to respectively $p_t\gtrsim2.5$ in
the central region $y_h=0,1$ where the DHJ curves deviate from the
data, and the entire forward region where both models work. We note
that a recent study of RHIC data on the nuclear suppression factor
$R^{dAu}$ suggests that the DHJ model already breaks down at $y_h=3.2$
\cite{Betemps:2008yw}.

%%%%%%%%%%%%%%%%%%
%%%%%%%%%%%%%%%%%%%%

\begin{figure}[htb]
\centering
\vspace{-15pt}
\includegraphics*[width=75mm]{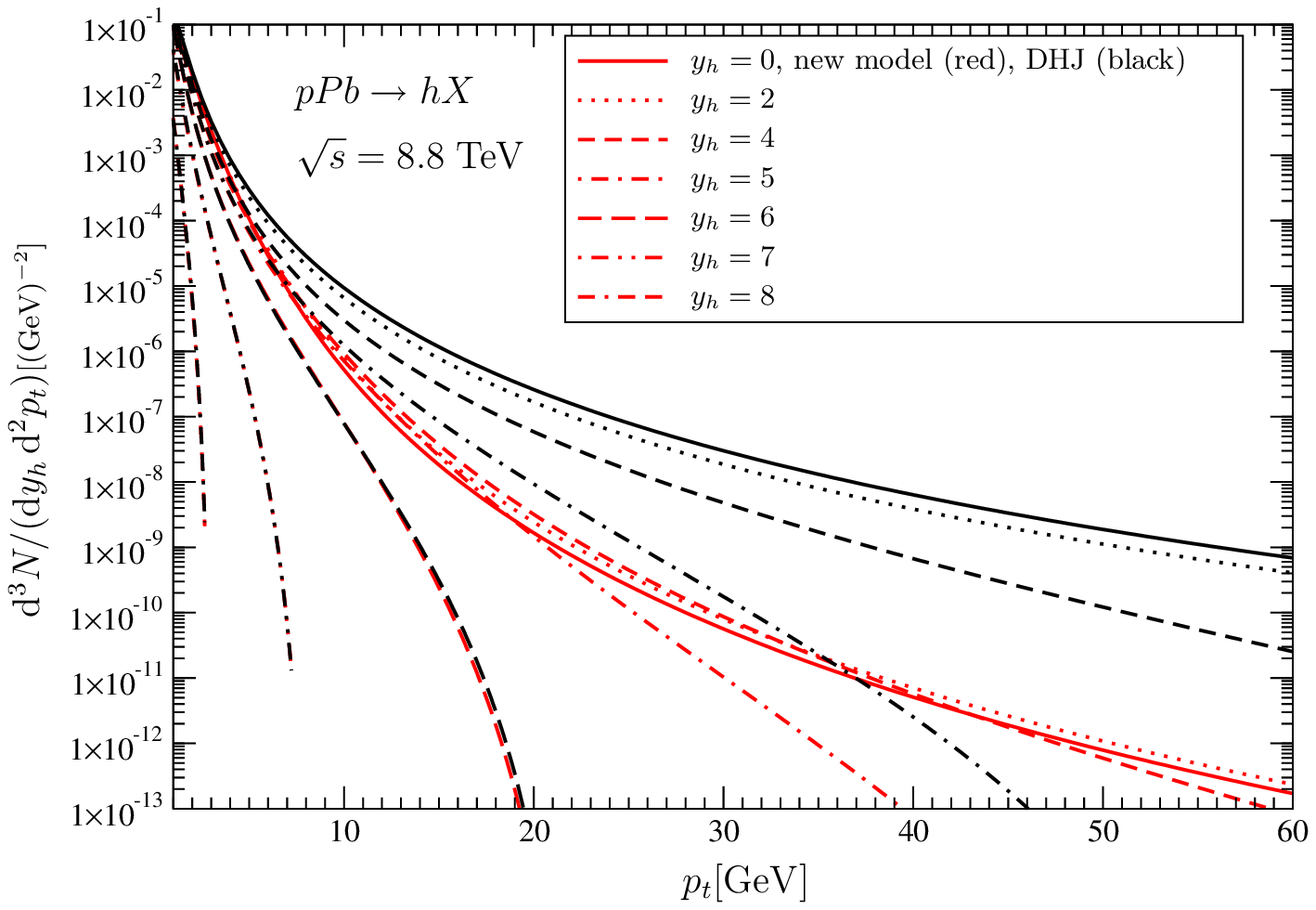}
\vspace{-35pt}
\caption{\label{fig_lhc} Hadron production cross section in $p\,$-$Pb$
  collisions at LHC from $\gamma_{\rm GS}$ and $\gamma_{\rm DHJ}$ at
  different rapidities $y_h$.}
\vspace{-5pt}
\end{figure}

Where the DHJ model curves deviate from the RHIC data, the probed
$x_2$ values are not very small and one may argue that a small-$x$
description cannot be expected to apply in the first place. However,
at LHC, due to the higher energies, the region of small $x_2$ extends
to a much larger range of $p_t$, so that the predictions of the DHJ
model and the new one will be different even at small $x_2$.
Fig.~\ref{fig_lhc} shows the $p_t$ distribution of hadron production
in $p\,$-$Pb$ scattering at 8.8 TeV at LHC. We emphasize again that
the prediction of the $p_t$ distribution using the GS model is to be
considered a tool for checking whether certain small-$x$ properties
are present in the data.  For the scaling curves the best fit obtained
from the RHIC data was used.

For small $p_t$ the predictions of the two models are comparable, since
there only the region of small values of $w$ is probed where
$\gamma_{\rm DHJ}$ and $\gamma_{\rm GS}$ are similar, cf.\
Fig.~\ref{fig_gam}. Also in the very forward region, i.e.\ $y_h
\approx 6 - 8$, only this region is tested and one obtains similar
results from both models.

However, there is quite a large range where the probed values of $x_2$
are small but the predictions are clearly different. From
Fig.~\ref{fig_lhc} we see that this is the case for rapidities smaller
than about 5 and $p_t\gtrsim5$ GeV---this region corresponds to
approximately $w\gtrsim 3$, where according to Fig.~\ref{fig_gam} the
DHJ model indeed deviates from the experimentally allowed band.  Due
to the larger energy, the values of $x_2$ at LHC are much smaller than
in the corresponding region of $p_t\gtrsim2.5$ GeV at RHIC. At LHC,
$x_2$ remains below 0.001 in the entire range depicted in
Fig.~\ref{fig_lhc}, i.e.\ even in the central region. Hence, if the
LHC data turns out to be described by the GS model, one would have a
clear indication of consistency with $d\,$-$Au$ data from RHIC at
forward {\em and} mid-rapidity within the small-$x$ framework of Eq.\
(\ref{eq:conv2}).

The main difference between the GS model and the DHJ model is the
slope of the resulting $p_t$ distribution, which is directly related
to the rise of $\gamma$ towards 1. Hence, a measurement of the slopes
of the $p_t$ distribution at moderate rapidities $y_h$ at LHC allows a
discrimination between the two models in a region where small-$x$
physics---and hence a description in terms of
Eq.~(\ref{eq:conv2})---is expected to be applicable. The slower
fall-off of the $p_t$ distribution in the DHJ model is caused by the
logarithmic rise of $\gamma$ towards 1, which is a generic signature
of small-$x$ evolution. Hence, if the $p_t$ distribution of the
produced hadrons falls off much faster than predicted by the DHJ
model, current expectations of small $x$ evolution do not hold at LHC.

%%%%%%%%%%%%%%%%%%
%%%%%%%%%%%%%%%%%%%%

%%%%%%%%%%%%%%%%%%%%%%%%%%%%%%%%%%%%%%%%%%%%%%%%%%%%%%%%%%%%%%%%%%%%%%%%%

\end{document}